# Growth of Large Domain Epitaxial Graphene on the C-Face of SiC

Rui Zhang[1], Yunliang Dong[2], Wenjie Kong[2], Wenpeng Han[3], Pingheng Tan[3], Zhimin Liao[1], Xiaosong Wu[1a)], and Dapeng Yu[1]

[1]State Key Laboratory for Artificial Microstructure and Mesoscopic Physics, Peking University, Beijing 100871, China

[2]Department of Physics, Guangxi Normal University, Guilin, Guangxi 541004, China

[3]State Key Laboratory for Superlattices and Microstructures, Institute of Semiconductors, Chinese Academy of Sciences, Beijing 100083, China

Growth of epitaxial graphene on the C-face of SiC has been investigated. Using a confinement controlled sublimation (CCS) method, we have achieved well controlled growth and been able to observe propagation of uniform monolayer graphene. Surface patterns uncover two important aspects of the growth, i.e. carbon diffusion and stoichiometric requirement. Moreover, a new "stepdown" growth mode has been discovered. *Via* this mode, monolayer graphene domains can have an area of hundreds of square micrometers, while, most importantly, step bunching is avoided and the initial uniformly stepped SiC surface is preserved. The stepdown growth provides a possible route towards uniform epitaxial graphene in wafer size without compromising the initial flat surface morphology of SiC.

a) Electronic mail: xswu@pku.edu.cn

# I. INTRODUCTION

Electronic applications of graphene require uniform film with high mobility in wafer size. Epitaxial growth seems a viable way to prepare the material that can meet these requirements. Epitaxial graphene (EG) on SiC shows great promise because of its high quality and potential to be grown in large size.[1,2] In particular, EG on the C-face of SiC, can have a mobility significantly higher than the one on the Si-face.[3,4] Although high speed transistors with a cut-off frequency up to 100 GHz, as well as integrated circuit fabrication, have already been demonstrated on EG on SiC,[5] the quality of the material is still far from reaching its potential. On the Si-face, graphene films often vary in thickness by a few layers.[6] Even for films grown in Argon, in which monolayer graphene dominates, thicker layers are found at step edges.[2,7] On the C-surface of SiC, the thickness variation is even larger due to a much faster growth rate.[3,8] Another issue is the domain size for graphene on SiC. It is generally believed that the growth of graphene starts at step edges and hence the graphene nucleation density is high.[9,10-12] Coalescence of islands can lead to domain boundaries at step edges, especially for bunched step edges. High resolution transmission electron microscopy has in fact shown that graphene films at these edges are defective.[12] Film quality can be markedly improved by increasing the domain size, as already been demonstrated in chemical vapour deposited graphene on metal.[13] On the other hand, studying the nucleation and propagation of graphene domains can provide insights into growth mechanism. However, such study turned to be difficult on the C-face. The growth on this surface is fast. The surface undergoes dramatic changes. Significant step bunching occurs on the SiC substrate, forming irregular large steps and the initial stepped surface almost completely disappears.[8] Consequently, the growth of the C-face graphene and the evolution of the substrate surface can hardly be tracked, which essentially hinders us from understanding the growth kinetics. Although investigation of the formation of graphene domains has previously been carried out and graphene islands were observed, the surface morphology displayed discontinuity at the edge of the island.[14] Also, step bunching is pronounced. There is still little known about how the SiC step decomposes and graphene nucleates and propagates on this surface.

In this work, we grow EG on SiC using the confinement controlled sublimation (CCS) method developed by de Heer team.[15] This method substantially increases the local Si pressure over the substrate. The growth consequently takes place at a slow rate in a condition close to thermodynamic equilibrium.[16] Thus, we are able to grow uniform graphene islands. Particularly, the initial half-unit-cell high SiC steps are preserved underneath the graphene film and continuous at the boundary of the islands. Several universal features unveil two salient factors that affect the growth, e.g. carbon diffusion and stoichiometry. Furthermore, a "stepdown" growth has been identified and counter-intuitively, it is favored against the "climbover" growth in our growth condition.[17] While the climbover growth tends to create step bunching, which roughens the surface, the stepdown growth keeps the regular SiC steps almost intact. Our results present reproducible and well-defined growth features that theoretical models can be compared with, which are relatively lacked so far. Furthermore, the stepdown process offers a method to realize uniform growth of large size graphene without degrading the initially flat morphology of the SiC surface.

## II. EXPERIMENTAL DETAILS

Growth of EG has been investigated on both on-Axis 6H- and 4H-SiC semi-insulating substrates, which were purchased from Tankeblue and Cree, respectively. Prior to growth, the SiC substrates were hydrogen etched at 1600 ºC so as to obtain atomically flat (but stepped due to a miscut angle) surface. The growth was carried out in a home-made high vacuum induction furnace. The SiC chips were placed in a graphite enclosure, which was carefully designed such that it is sealed as tight as possible. The design provides the essential confinement for silicon sublimation. The pressure in the vacuum furnace was around 1×10-4 Torr during growth. The temperature of the sample was measured by a type C thermal couple. In order not to disturb the growth and break the seal, the thermocouple was placed in an enclosure that is in a position symmetric to the growth enclosure and therefore is assumed to have the same temperature as the sample. Samples were first annealed at 1000 ºC for 20 min to remove the native oxide on the surface. They were then heated to the growth temperature of 1560 ºC. After 10-20 min of growth, heating was shut off and the samples were allowed to cool naturally. Atomic force microscopy (AFM) in a tapping mode was employed and both the topography and phase data were collected and analysed. Raman measurements were performed in a backscattering geometry using a Jobin-Yvon HR800 Raman system equipped with a liquid nitrogen cooled charge-coupled detector. The laser excitation wavelength is 532 nm from a diode-pumped solid-state laser.

## III. RESULTS AND DISCUSSION

The hydrogen-etched SiC surface exhibits terraces of about equal width, as shown in Fig. 1(a). The steps are full-unit-cell high, 1.5 nm for 6H-SiC and 1 nm for 4H-SiC. After 12 min CCS growth at 1560 ºC, each step splits into two half-unit-cell high steps, seen in Figs. 1(b) and 1(c). The nearly periodic steps are attributed to a self-ordering process, which suggests that the surface evolution takes place in a condition close to equilibrium.[18] Such condition is essential for growth of high quality graphene. The change of the step morphology results from a dynamic balance between decomposition and recombination of SiC during annealing and redistribution of chemical elements on the surface. It is inferred that carbon can diffuse from one step edge to another at 1560 ºC. A similar carbon diffusion ability, indicated by significant step bunching, has been observed on Si-face SiC annealed in Argon.[2,7] We will show below that the substantial diffusivity of carbon is crucial for growth of uniform graphene.

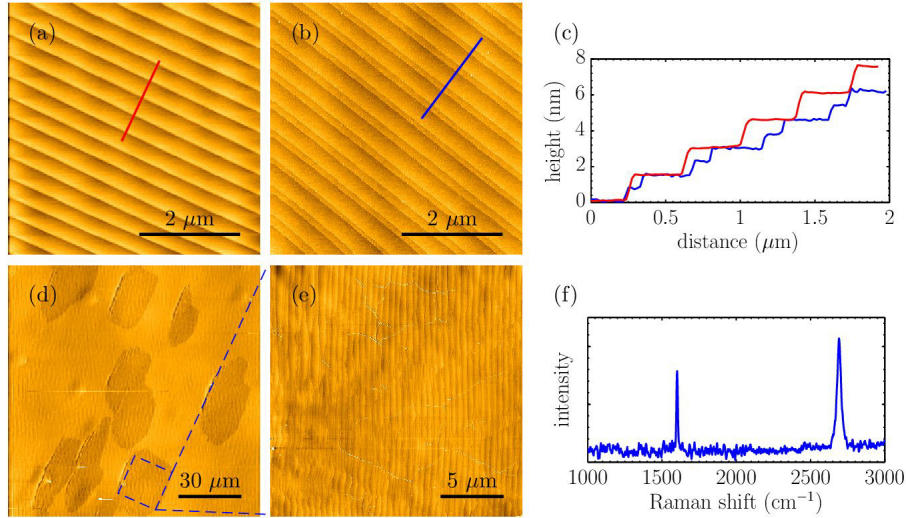

FIG. 1. Growth of graphene islands on the C-face of 6H-SiC substrate. (a) AFM image of the SiC surface prior to graphitization showing evenly spaced steps. (b) AFM image of the SiC surface that is not covered by graphene after growth at 1560 ºC for 12 min. (c) AFM line profiles for the colored lines in (a) and (b) showing that the step heights are 1.5 nm (6 SiC bilayers) and 0.75 nm (3 SiC bilayers) prior to and after annealing, respectively. Before extracting the profiles, the images were leveled such that each terrace appeared at the same height. (d) Large-scan AFM image of graphene islands on the surface. (e) Zoom-up AFM image for the area indicated by the blue dashed square in (d). The image shows graphene with pleats on the uniform SiC steps. (f) Raman spectrum for a graphene island. SiC background has been subtracted. The spot size of the laser excitation is about 2 micrometer.

Besides the morphological change of SiC steps, graphene islands form on the surface. As seen in Fig. 1(d), the islands appear as slight depressions on the surface. Pleats, also called ripples, puckers or ridges, are due to the different thermal expansion coefficients and the weak coupling between SiC and graphene.[19] They are characteristic of the C-face graphene and hence serve as a faithful indication for graphene. Figure 1(e) shows a zoom-up image for an island in Fig. 1(d). Pleats, manifested as faint white lines in the image, are indeed present in the depressed area. Note that the phase signal of a tapping mode AFM can well distinguish SiC and graphene.[20,21] We have seen a strong contrast in the phase image for this area, which confirms that the depressions are covered by graphene. Similar phase image will be shown later. A typical

Raman spectrum, plotted in Fig. 1(f), shows the characteristic of graphene. The absence of the D peak suggests the high quality of EG. The G and 2D peaks locate at 1602 cm$^{-1}$ and 2692 cm$^{-1}$, respectively, which are 20 cm$^{-1}$ and 17 cm$^{-1}$ higher than the corresponding peaks in exfoliated graphene at the same excitation energy.[22] The higher mode frequency in EG results from the compressive stress in the graphene layers on SiC substrate.[23] All islands are elongated along the step, suggesting that the barrier for propagation of graphene is smaller in this direction. The morphology of the graphene island is highly uniform. No apparent step bunching occurs. The size of the island is usually about a few hundred square micrometers and can be up to a thousand, as seen in Fig. S-1 in supplementary material at [URL will be inserted by AIP] for optical image of the C-face graphene on 6H-SiC. Note that another group has recently reported a monolayer graphene island of two hundred square micrometers on non-uniform SiC steps.[24] Growth of large islands originates from a nucleation energy being large than the propagation energy.[17] We have carried out growth for different periods of time and found that the shorter the annealing time, the lower the island density is and the smaller the islands are.

To see the details of an island, we show AFM images of a smaller island in Fig. 2. More examples of the islands can be found in Fig. S-2 and S-3 in supplementary material at [URL will be inserted by AIP] for AFM and DFM images of graphene islands. An interesting feature is that the graphene island is bounded by a halo of particles. There are three possibilities for what these particles consist of, e.g. silicon, carbon and SiC. We argue in the following that these are indeed carbon particles. Since the growth temperature is much higher than the melting point of silicon, especially silicon particles in nanometer size, we can safely rule it out. These are unlikely SiC particles, either, because the step flow on the SiC surface suggests that SiC particles tend to stay at the step edge and crystallize to minimize free energy. Therefore, these particles are made of carbon. When SiC decomposes, carbon atoms are liberated and diffuse on the surface. The width of the carbon particle halo, about 1 μm, suggests that carbon atoms can travel micron distances, in agreement with the substantial carbon diffusivity implied by the morphological change of SiC.

Furthermore, the change of the step morphology due to growth is conspicuous, as shown in Fig. 2. In particular, the steps bend in the step up direction when they approach the graphene island and become wavy underneath it. These features are apparently the result of the step recession, which can be understood based on a simple consideration for the stoichiometric requirement. Since only steps of three SiC bilayers high have been found on graphene islands, it is reasonable to assume that only integer numbers of SiC triple bilayers were decomposed. The step edge recedes towards the upper terrace upon decomposition of a terrace. If three bilayers of SiC are uniformly removed from the surface by decomposition, step edges will remain at their original positions. However, considering the carbon areal density, 3.14 SiC bilayers are necessary to decompose to form a monolayer graphene. So, more SiC triple bilayers are required. This leads to further recession of step edges with respect to their original position, which is responsible for the observed morphological change of steps. Complex step morphology due to stoichiometry and carbon diffusion has been observed when only a single SiC bilayer was decomposed.[11]

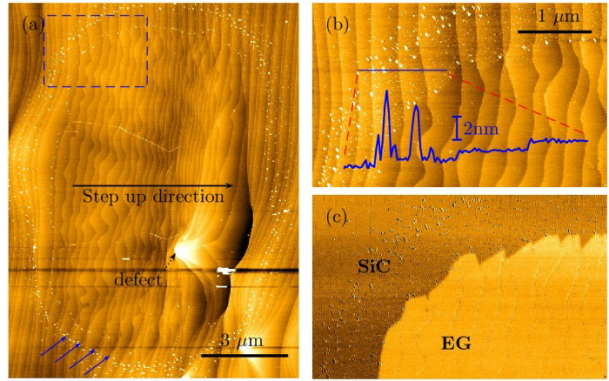

FIG. 2. Nucleation of graphene. (a) AFM image of a graphene island grown at 1560 ºC for 10 min. A halo of carbon particles circumscribes the island. A defect, manifested as a large particle, is found inside of the island, but closer to the upper edge (the right side of the image), where significant step bunching appears. Step recession is seen at the proximity of the island edge, indicated by blue arrows. (b) Close-up AFM image of the area enclosed by the dashed blue rectangle in (a). Graphene is recognized by its smoother morphology. A line profile is plotted for the thin blue line. (c) Phase channel of the close-up image highlights the graphene island.

In Fig. 2(a), a particle, most likely manifest of a SiC defect, appears inside the graphene island. Such a defect is seen in most of graphene islands, strongly suggesting that it is a nucleation center and graphene is a single domain. In the CCS method, the Si pressure over the surface is relatively high so that the sublimated silicon atoms return to the surface and recombine into SiC with carbon atoms. In this regard, silicon acts as a carbon etchant. The etching effect strongly suppresses nucleation of graphene. As a consequence, the nucleation energy in the CCS method is high. Graphene islands are hard to nucleate unless a defect is present. This is consistent with previous experiments on the early stage of graphene growth.[14] On the other hand, in a UHV condition, the sublimated silicon never comes back to the surface. Abundant carbon atoms left on the surface facilitate the graphene nucleation, resulting in dense and small islands.[21] The coalescence of the islands likely creates a large amount of grain boundaries. So, to obtain high quality EG, it is necessary to promote island expansion instead of nucleation of more islands. This means the propagation energy should be smaller than the nucleation energy.[17]

Severe step bunching is found on the upper side of all graphene islands. Additionally,

nucleation centers are always closer to the step bunch, especially when the distance is measured in terms of the number of steps. In the step up direction, graphene is only able to climb over one or two steps and then stops at a step bunch. On the other hand, graphene propagates over many steps in the step down direction. Two growth modes have previously been proposed for graphene growth on SiC.[17] In a coalescence type of growth, graphene nucleates at step edges and coalesces when it meets the graphene film on the upper terrace. When the nucleation density is low, a so-called "climbover" growth can take place. In this growth mode, when graphene propagates to the step edge of the upper terrace, it climbs over the step and continue to grow on the upper terrace. However, our experiment reveals a new type of growth, in which graphene grows in the step down direction. Figure 3 shows an AFM image for the front of the stepdown growth. After a terrace is decomposed and subsequently replaced by graphene, the graphene growth front lies in the middle of a flat terrace. But, the growth doesn't stop here; instead, SiC right under the growth front starts to decompose. As a result, the graphene film steps down to the lower terrace and continues to propagate. We call the process "stepdown" growth. Step edges are usually considered as less stable and decompose first.[9,12] In the stepdown growth, it is not clear why decomposition takes place in the middle of the terrace. One possibility is that the heat released by crystallization of carbon atoms into graphene produces an increase in local temperature,[25] which facilitates decomposition. One may think that there is a significant higher energy barrier for the stepdown process than for the propagation on a terrace. However, the fact that most islands only exhibit slight elongation along steps implies that the propagation speed in the step down direction isn't significantly lower (See islands in Fig. 2(a) and Fig. S-3 in supplementary material at [URL will be inserted by AIP] for AFM images of graphene islands). This is in sharp contrast to a previous experiment where extremely long graphene islands were found.[26]

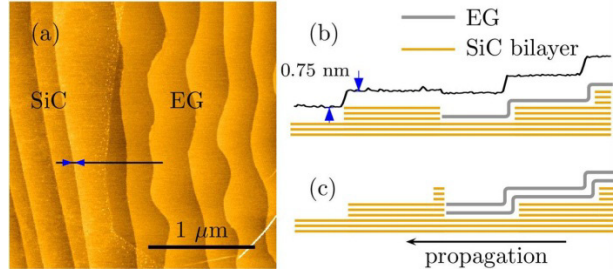

FIG. 3. Stepdown growth. (a) AFM image of a graphene edge at the lower side of an island, showing graphene propagates in the step down direction. (b) The morphological profile for the thin black line in (a). The image was leveled before extraction of the profile. Steps on both SiC and graphene are three SiC bilayers high, 0.75 nm. The blue arrows in (a) and (b) indicates where the height is measured. A schematic model is proposed to account for the line profile. (c) An alternative model for propagation of bilayer graphene. Once the bilayer graphene film reaches the step edge, the propagation of the top graphene layer stops, while the bottom one continues to grow, just like the monolayer case in (b).

As showing in Fig. 3, the zoom-up AFM image at the island boundary doesn't display a gap between the graphene island and the bare SiC surface, indicating that the exposed SiC surface after decomposition is immediately covered by newly formed graphene. As we discussed above, three bilayers cannot provide sufficient carbon for a graphene layer. Therefore, the rest of carbon has to come from decomposition of neighbor steps. The long diffusion length of carbon ensures that stoichiometry can be satisfied in this way. Such carbon transfer is responsible for the recession of the neighbor step edges towards the upper terraces, which causes step bunching in the step up direction, but not in the step down direction. Step bunching prevents graphene from propagating in the step up direction. The importance of the carbon diffusivity can be further elaborated by considering a very low carbon diffusivity. In such case, no sufficient carbon can be supplied to the growth front, causing separation of the growth front from the decomposition front. Once the separation becomes larger than the diffusion length, graphene propagation stops.

Graphene islands most likely consist of only one layer of graphene, because the stepdown growth is not applicable to the propagation of bilayer graphene, which involves decomposition of two SiC triple bilayers simultaneously. As illustrated in Fig. 3(c), if a bilayer graphene film propagates to a step edge, there would still be a SiC triple bilayer on the same level with the bottom graphene layer. As a consequence, the bottom layer would continue to propagate by decomposition of this SiC triple bilayer. But, the propagation of the top layer would stop. Optical microscope images show that most of islands have the same and faintest contrast, indicating a monolayer graphene film. In Fig. 3(b), the height difference between the SiC surface and the graphene film at the growth front is 0.2±0.1 nm, close to 0.4 nm (0.75 nm - 0.34 nm) for the case of a monolayer graphene film, corroborating with the optical measurements. However, caution need to be taken when interpreting the height difference between graphene and other surfaces.[27]

We summarize the graphene growth mechanism in our experiment in Fig. 4. The nucleation energy is high due to high local Si pressure. Thus, nucleation often occurs at a defect instead of a step edge. Because of stoichiometry, part of carbon for growth comes from decomposition of the neighbor steps, which leads to recession of these step edges in the step up direction. The recession generates step bunching at the upper edge of the island, which stops the propagation of graphene. On the other hand, the recession suppresses step bunching in the step down direction by increasing the distance between the steps on the two sides of the growth front. By the unexpected stepdown process, graphene propagates in the step down direction. The stepdown growth represents a sustainable expansion of graphene and can be used to grow uniform graphene in large size.

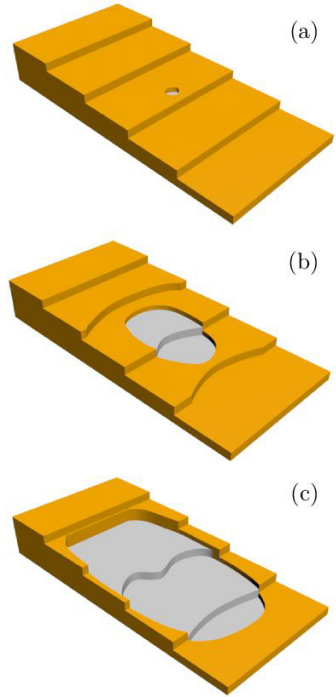

FIG. 4. A schematic diagram for the graphene growth on the C-face of SiC. Dark yellow represents bare SiC surface, while gray represents graphene. (a) Graphene nucleation. (b) The propagation of graphene in the step up direction and the step down direction, inducing recession of the neighbor steps. (c) The recession of the neighbor step causes step bunching at the upper side, which prevents graphene from propagating, while graphene continues propagating in the step down direction.

On 4H-SiC, the half-unit-cell step is two bilayers high. Since 3.14 bilayers are necessary for formation of a monolayer graphene, it has been suspected that 4H-SiC is less suitable for graphene growth than 6H-SiC. We have carried out experiments on 4H-SiC. Similar nucleation of graphene islands has also been observed under the same growth condition. Most features, e.g., splitting of SiC steps, the carbon particle halo and recession of SiC steps near graphene edges, are observed, which suggests that the growth mechanism discussed above is applicable to 4H-SiC, too. Nevertheless, a few differences exist. First, though the step height is two bilayers on the bare SiC surface, it becomes four bilayers high underneath the graphene island. The obvious consequence of disappearance of every other terrace is that more carbon atoms are liberated. The abundance of carbon atoms may explain the second difference that step edges are much more straight on 4H-SiC than on 6H-SiC. In fact, a straight and nearly periodically stepped surface is preserved after growth, shown in Fig. 5(b). We want to point out that it perhaps represents the ideal graphene on SiC in terms of morphology.

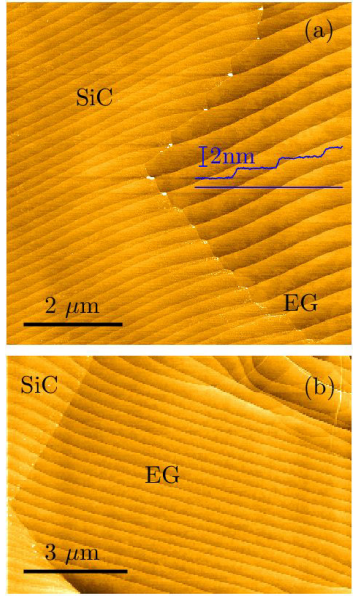

FIG. 5. AFM images of graphene grown on the C-face of 4H-SiC at 1560 ℃ for 12 min. (a)AFM image taken at the edge of a graphene island. Going from the bare SiC surface to the graphene island, two half-unit-cell high steps bunch into a full-unit-cell high step. Carbon particles are also found on SiC near the edge. A line profile is shown on top of the image. (b) Graphene film on evenly spaced SiC steps.

## IV. CONCLUSIONS

We have achieved highly controllable growth of EG on the C-face of 6H- and 4H-SiC. Uniform monolayer, single domain graphene islands up to a few tens of micrometers in lateral size have been grown. The morphological changes that the surface undergoes reveal a substantial diffusivity of carbon. The carbon diffusion not only explains how the stoichiometry requirement is met during growth, but also plays an important role in shaping the surface morphology and maintaining the graphene propagation. Although two processes, coalescence and climbover, were previously believed to account for the growth of graphene on the Si-face, they are not favored in our experimental condition. Instead, a unexpected "stepdown" process has been identified for the first time. It is the dominant growth mode and promotes propagation of graphene perpendicular to SiC steps without causing step bunching. Thus, it opens a route towards growth of uniform graphene in wafer size on the C-face SiC.

## ACKNOWLEDGEMENTS

This work was supported by National Science Foundation of China (NSFC project 11074007) and Ministry of Science and Technology (MOST No. 2012CB933404, 2012CB933401, 2009CB623703). We also acknowledge the International Science Technology Cooperation Program of China Sino Swiss Science and Technology Cooperation Program (SSSTC, 2010DFA01810).

## REFERENCES

[1] C. Berger, Z. M. Song, T. B. Li, X. B. Li, A. Y. Ogbazghi, R. Feng, Z. T. Dai, A. N. Marchenkov, E. H. Conrad, P. N. First, and W. A. de Heer, J. Phys. Chem. B **108** (52), 19912 (2004); C. Berger, Z. M. Song, X. B. Li, X. S. Wu, N. Brown, C. Naud, D. Mayo, T. B. Li, J. Hass, A. N. Marchenkov, E. H. Conrad, P. N. First, and W. A. de Heer, Science **312** (5777), 1191 (2006).

[2] Konstantin V. Emtsev, Aaron Bostwick, Karsten Horn, Johannes Jobst, Gary L. Kellogg, Lothar Ley, Jessica L. McChesney, Taisuke Ohta, Sergey A. Reshanov, Jonas Rohrl, Eli Rotenberg, Andreas K. Schmid, Daniel Waldmann, Heiko B. Weber, and Thomas Seyller, Nat. Mater. **8** (3), 203 (2009).

[3] W. A. de Heer, C. Berger, X. S. Wu, P. N. First, E. H. Conrad, X. B. Li, T. B. Li, M. Sprinkle, J. Hass, M. L. Sadowski, M. Potemski, and G. Martinez, Solid State Commun. **143** (1-2), 92 (2007).

[4] Xiao Song Wu, Xue Bin Li, Zhi Min Song, Claire Berger, and Walt A. de Heer, Phys. Rev. Lett. **98**, 136801 (2007); Xiao Song Wu, Yike Hu, Ming Ruan, Nerasoa K. Madiomanana, John Hankinson, Mike Sprinkle, Claire Berger, and Walt A. de Heer, Appl. Phys. Lett. **95** (22), 223108 (2009).

[5] Y. M. Lin, C. Dimitrakopoulos, K. A. Jenkins, D. B. Farmer, H. Y. Chiu, A. Grill, and Ph Avouris, Science **327** (5966), 662 (2010); Y. M. Lin, A. Valdes-Garcia, S. J. Han, D. B. Farmer, I. Meric, Y. N. Sun, Y. Q. Wu, C. Dimitrakopoulos, A. Grill, P. Avouris, and K. A. Jenkins, Science **332** (6035), 1294 (2011).

[6] H. Huang, W. Chen, S. Chen, and A. T. S. Wee, Acs Nano **2** (12), 2513 (2008).

[7] C. Virojanadara, M. Syväjarvi, R. Yakimova, L. I. Johansson, A. A. Zakharov, and T. Balasubramanian, Phys. Rev. B **78** (24), 245403 (2008).

[8] Luxmi, N. Srivastava, G. He, R. M. Feenstra, and P. J. Fisher, Phys. Rev. B **82** (23), 235406 (2010).

[9] J. B. Hannon and R. M. Tromp,   Phys. Rev. B **77** (24), 241404 (2008);   Siew   Wai Poon, Wei Chen, Eng Soon Tok, and Andrew T. S. Wee,   Appl. Phys. Lett. **92** (10), 104102 (2008).

[10] W. Norimatsu and M. Kusunoki,   Physica E **42** (4), 691 (2010).

[11] Taisuke Ohta, N. C. Bartelt, Shu Nie, Konrad Thürmer, and G. L. Kellogg,   Phys. Rev. B **81** (12), 121411 (2010).

[12] J. Robinson, X. J. Weng, K. Trumbull, R. Cavalero, M. Wetherington, E. Frantz, M. LaBella, Z. Hughes, M. Fanton, and D. Snyder,   Acs Nano **4** (1), 153 (2010).

[13] Xuesong Li, Carl W. Magnuson, Archana Venugopal, Jinho An, Ji Won Suk, Boyang Han, Mark Borysiak, Weiwei Cai, Aruna Velamakanni, Yanwu Zhu, Lianfeng Fu, Eric M. Vogel, Edgar Voelkl, Luigi Colombo, and Rodney S. Ruoff,   Nano Lett. **10** (11), 4328 (2010).

[14] N. Camara, G. Rius, J. R. Huntzinger, A. Tiberj, L. Magaud, N. Mestres, P. Godignon, and J. Camassel,   Appl. Phys. Lett. **93** (26), 263102 (2008);   J. L. Tedesco, G. G. Jernigan, J. C. Culbertson, J. K. Hite, Y. Yang, K. M. Daniels, R. L. Myers-Ward, C. R. Eddy, Jr., J. A. Robinson, K. A. Trumbull, M. T. Wetherington, P. M. Campbell, and D. K. Gaskill,   Appl. Phys. Lett. **96** (22), 222103 (2010);   J. K. Hite, M. E. Twigg, J. L. Tedesco, A. L. Friedman, R. L. Myers-Ward, C. R. Eddy, and D. K. Gaskill,   Nano Lett. **11** (3), 1190 (2011).

[15] Walt A. de Heer, Claire Berger, Ming Ruan, Mike Sprinkle, Xuebin Li, Yike Hu, Baiqian Zhang, John Hankinson, and Edward Conrad,   PNAS **108**, 16900 (2011).

[16] R. M. Tromp and J. B. Hannon,   Phys. Rev. Lett. **102** (10), 106104 (2009).

[17] Fan Ming and Andrew Zangwill,   Phys. Rev. B **84**, 115459 (2011).

[18] Hiroshi Nakagawa, Satoru Tanaka, and Ikuo Suemune,   Phys. Rev. Lett. **91**, 226107 (2003).

[19] J. Rohrl, M. Hundhausen, K. V. Emtsev, T. Seyller, R. Graupner, and L. Ley,   Appl. Phys. Lett. **92** (20), 201918 (2008).

[20] Michael L. Bolen, Sara E. Harrison, Laura B. Biedermann, and Michael A. Capano,   Phys. Rev. B **80** (11), 115433 (2009).

[21] F. J. Ferrer, E. Moreau, D. Vignaud, D. Deresmes, S. Godey, and X. Wallart,   J. Appl. Phys. **109** (5), 054307 (2011).

[22] W. J. Zhao, P. H. Tan, J. Liu, and A. C. Ferrari,   Journal of the American Chemical Society **133** (15), 5941 (2011).

[23] Z. H. Ni, W. Chen, X. F. Fan, J. L. Kuo, T. Yu, A. T. S. Wee, and Z. X. Shen,   Phys. Rev. B **77** (11), 115416 (2008).

[24] Yike Hu, Ming Ruan, Zelei Guo, Rui Dong, James Palmer, John Hankinson, Claire Berger, and Walt A. de Heer,   Journal of Physics D: Applied Physics **45** (15), 154010 (2012).

[25] Valery Borovikov and Andrew Zangwill,   Phys. Rev. B **80** (12), 121406 (2009).

[26] Nicolas Camara, Jean-Roch Huntzinger, Gemma Rius, Antoine Tiberj, Narcis Mestres, Francesc Pérez-Murano, Philippe Godignon, and Jean Camassel,   Phys. Rev. B **80** (12),

125410 (2009).

[27] P. Nemes-Incze, Z. Osváth, K. Kamarás, and L. P. Biró, Carbon **46** (11), 1435 (2008).

**Supplementary materials**

# Growth of Large Domain Epitaxial Graphene on the C-Face of SiC

Rui Zhang[1], Yunliang Dong[2], Wenjie Kong[2], Wenpeng Han[3], Pingheng Tan[3], Zhimin Liao[1],

Xiaosong Wu[1a)], and Dapeng Yu[1]

[1]*State Key Laboratory for Artificial Microstructure and Mesoscopic Physics, Peking University, Beijing 100871, China*

[2]*Department of Physics, Guangxi Normal University, Guilin, Guangxi 541004, China*

[3]*State Key Laboratory for Superlattices and Microstructures, Institute of Semiconductors, Chinese Academy of Sciences, Beijing 100083, China*

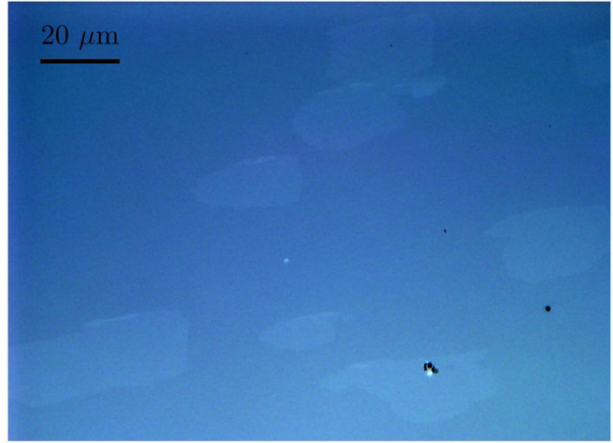

FIG. S-1. Optical image of the C-face graphene on 6H-SiC. Graphene islands display uniform and the faintest contrast, suggesting a monolayer. Only close to the upper edge of islands, slightly brighter areas appear, probably indication of initiation of the second layer.

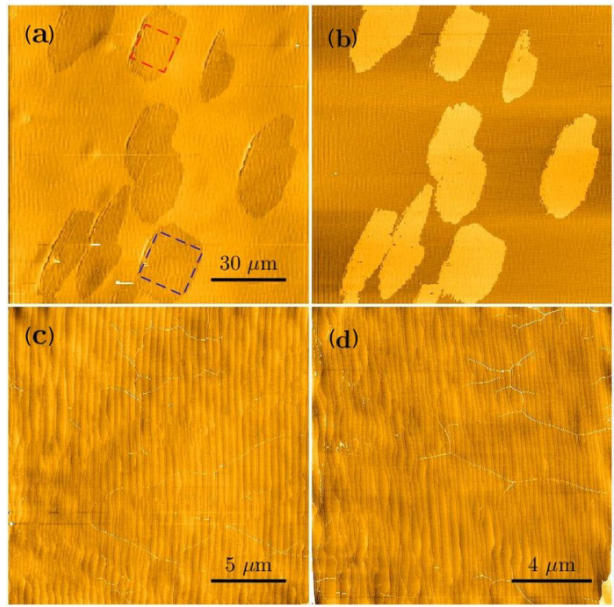

FIG. S-2. Graphene islands on the C-face of 6H-SiC substrate. (a) Figure 1(d) in the main text. (b) Corresponding phase image showing clear contrast between the graphene islands and the bare SiC surface. The phase signal makes identification of graphene easy and unambiguous. (c) Figure 1(e) in the main text. A zoom-up AFM topography image for the area marked by a dashed blue square in (a). (d)Zoom-up image of the area marked by a dashed red square in (a). Graphene with faint pleats covers the SiC surface, which is uniformly stepped. The step height is half-unit-cell, 0.75 nm.

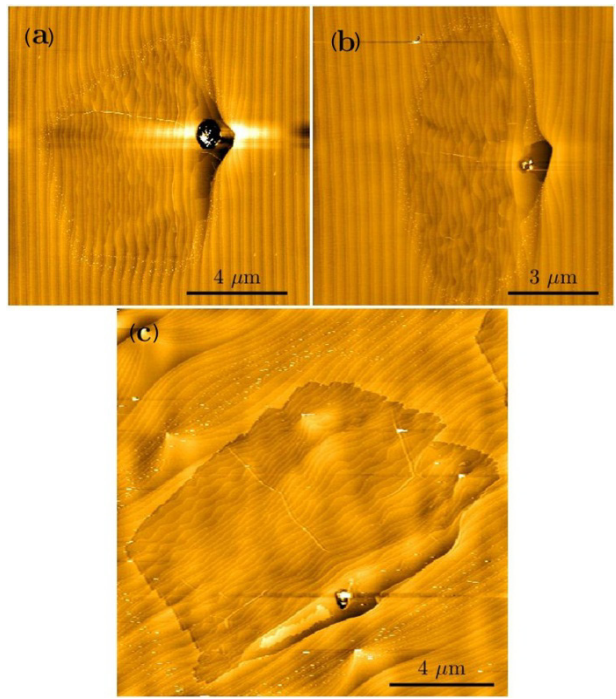

FIG. S-3. Three examples for nucleation and propagation of graphene on the C-face of 6H-SiC. All examples display the same features discussed in the main text. Islands are surrounded by a halo of carbon particles. Graphene nucleates from a SiC defect. The defect is closer to the upper edge of the island, at which severe step bunching occurs. SiC steps recede in the step up direction when approaching the island.